\documentclass[10.5pt , twocolumn]{article}
\usepackage{graphicx}
\usepackage{epstopdf}
\usepackage{dcolumn}
\usepackage{verbatim}
\usepackage{float}
\usepackage{slashed}
\usepackage{wrapfig}
\usepackage{subcaption}
\usepackage{hyperref}

\newcommand\pubnumber{DPF2017-10}
\newcommand\pubdate{\today}

\def\mit{Laboratory for Nuclear Science\\
Massachusetts Institute of Technology\\77 Massachusetts Avenue, Bldg 26-568, MA 02139, U.S.A}
\def\support1{}

\def\msstate{Department of Physics and Astronomy\\
Mississippi State University\\P. O. Box 5167, Mississippi State, MS 39762, U.S.A}
\def\support2{}

\def\Title#1{\begin{center} {\Large #1 } \end{center}}
\def\Author#1{\begin{center}{ \sc #1} \end{center}}
\def\Address#1{\begin{center}{ \it #1} \end{center}}

\newcommand\pubblock{\rightline{\begin{tabular}{l} \pubnumber\\
         \pubdate  \end{tabular}}}
\newenvironment{Abstract}{\begin{quotation}  }{\end{quotation}}
\newenvironment{Presented}{\begin{quotation} \begin{center} 
             PRESENTED AT\end{center}\bigskip 
      \begin{center}\begin{large}}{\end{large}\end{center} \end{quotation}}

\def\beq{\begin{equation}}
\def\eeq#1{\label{#1}\end{equation}}
\def\eeqn{\end{equation}}

\def\beqa{\begin{eqnarray}}
\def\eeqa#1{\label{#1}\end{eqnarray}}
\def\eeqan{\end{eqnarray}}

\let\bar=\overbar

\def\Dslash{\not{\hbox{\kern-4pt $D$}}}
\def\dslash{\not{\hbox{\kern-2pt $\del$}}}

\def\msb{{\bar{\ssstyle M \kern -1pt S}}}

\begin{document}
\begin{titlepage}
\pubblock

\vfill
\Title{CompEx II: A Pathway in Search of BSM Physics using Compton Scattering}
\vfill
\Author{Prajwal Mohanmurthy \footnote{prajwal@mohanmurthy.com . Also at ETH Zurich \& Paul Scherrer Institute.}}
\Address{\mit}
\Author{Dipangkar Dutta}
\Address{\msstate}
\vfill
\begin{Abstract}
Constancy and anisotropy of vacuum refractive index serves as a strong way to probe the predictions of theories beyond the standard model (BSM), especially those that predict breaking of local Lorentz and CPT symmetries. For photons of energies in the ranges $9-46$ MeV using a $1.16$ GeV electron beam, a constraint on vacuum refractive index, $(n-1) <  1.4 \times 10^{-8}$ was imposed using the Compton polarimeter in Hall - C of Jefferson Lab (JLab). Absence of sidereal modulation of the vacuum refractive index was then used to constrain the Minimal Standard Model Extension (MSME) parameters of $\sqrt{\kappa_X^2 + \kappa_Y^2} <  8.6 \times 10^{-10}$. These preliminary set of measurements will be followed up by measurements using the $11$ GeV electron beam at JLab with a sensitivity better than current leading constraints by a factor of $4-8$. Furthermore, quantum gravity models predict crystalline nature of space at Planck scales which may manifest as vacuum birefringence that can be probed by Compton scattering using circularly polarized light. We show that future facilities such as the ILC provide tantalizingly interesting possibilities.
\end{Abstract}
\vfill
\begin{Presented}
DPF 2017\\
The Meeting of the American Physical Society\\
Division of Particles and Fields\\
Fermilab, Batavia, IL, July 31 - August 4, 2017\\
\end{Presented}
\vfill
\end{titlepage}
\def\thefootnote{\fnsymbol{footnote}}
\setcounter{footnote}{0}

\section{Introduction}
\begin{figure*}[h!]
\centering
\includegraphics[scale=.45]{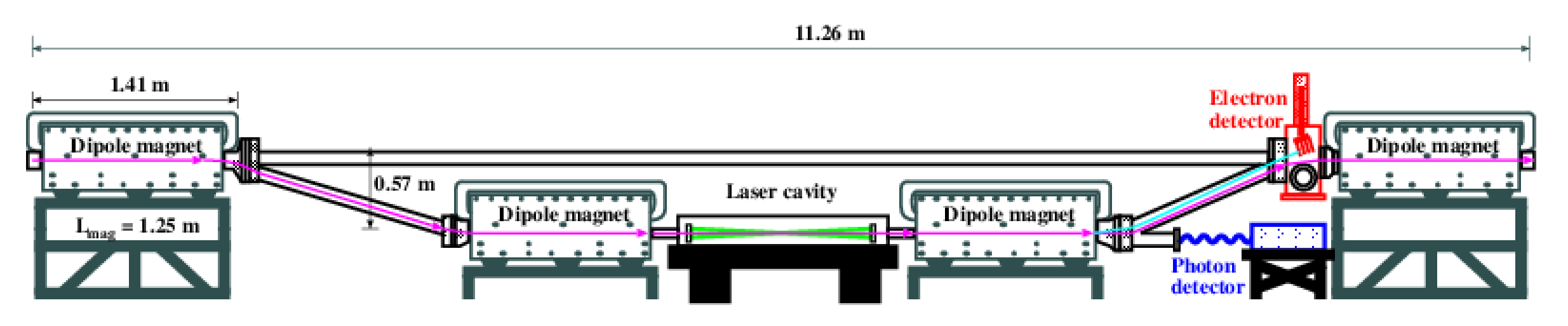}
   \caption[]{Schematic diagram of the JLab Hall-C Compton polarimeter, \textit{courtesy} Ref. \cite{[23]}.}
\label{fig1}
\end{figure*}
~~~Lorentz invariance (LI) is a common symmetry shared by both standard model (SM) and general relativity (GR). By Charge Conjugation-Parity-Time Reversal (CPT) theorem; violation of CPT symmetry implies a violation of local Lorentz symmetry \cite{[1],[2]}, meaning that any process that might break LI would indicate new physics beyond SM and GR \cite{[3]}. Planck-scale unified theories allow spontaneous violations of Lorentz symmetry \cite{[4],[5],[6],[7],[8]}, therefore searching for breakage of LI is a good test of Planck-scale physics \cite{[9]}. Even though no violation of Lorentz symmetry has ever been reported, it is not inconceivable to think of the presence of a Lorentz symmetry violating field in the universe as might be indicated by dark energy which dominates the universe \cite{[10],[11],[12],[13],[14]}. Lorentz symmetry guarantees the isotropy of speed of light and the independence of speed of light from the energy of the photons. Breakage of local LI is incorporated into an effective field theory at lab energies far below Planck-scale known as standard model extension (SME) \cite{[15],[16],[17]} which provides a framework to analyze experiments in search of Lorentz symmetry violation \cite{[18]}.

Interactions between SM particles and a LI breaking field would manifest themselves as violation of Lorentz or CPT symmetry \cite{[19]}. One such manifestation might be a preferred direction, $\hat{\kappa}$, in the universe which breaks LI and generates an anisotropy in the speed of light.
\begin{equation}
n(\hat{k}) \approx 1+\vec{\kappa}\cdot\hat{k}
\label{eq1}
\end{equation}
where $n(\hat{k})$ is the refractive index of free space in the direction of $\hat{k}$, and the components of the 3-vector $\tilde{\kappa}$ are the minimal SME parameters \cite{[20],[21]}. It is evident from Eq. \ref{eq1} that the measured speed of light is modulated sidereally as $\hat{\kappa}$ is held fixed in the cosmos whilst the direction-$\hat{k}$ goes with the rotating frame of the Earth. 

\section{CompEx-1a: Measurement in Hall-C of JLab}
A new polarimeter based on Compton scattering was built for the sub 5\% measurement of the weak charge of the proton at the QWeak experiment in Hall-C of Jefferson Lab (JLab) \cite{[22]}. The polarimeter is based on the measurement of the known double-spin Compton asymmetry in electron scattering from a photon beam of known polarization.
\begin{equation}
A_{exp} = \frac{\sigma^+ - \sigma^-}{\sigma^+ + \sigma^-}
\label{eq2}
\end{equation}
Here $\sigma^{+(-)}$ indicates the Compton cross-section for positive (negative) incident electron helicity while the incident photon helicity state is fixed. The polarimeter, as shown in Figure \ref{fig1}, consists of 4 dipole magnets, where the 3rd dipole magnet (from left) acts as an analyzing magnet for the Compton scattered electrons. The Compton scattering occurs in a low-gain Fabry-P\'erot cavity ($Q \in (100,300)$) coupled to a 10 W - 532 nm laser where the electrons had an initial energy of 1.16 GeV. The Compton scattered electrons are detected by a multi-strip diamond detector, and this allows us to extract Compton asymmetry (in Eq. \ref{eq3}) as a function of scattered electron energy. The Compton scattered photon had a maximum energy of up to 46 MeV and was detected by a calorimeter made of PbWO$_4$ scintillator crystals. Refs. \cite{[23],[24],[25]} provide further details of the construction and performance of the polarimeter. 

\begin{figure*}[h]
\centering
   \includegraphics[scale=.65]{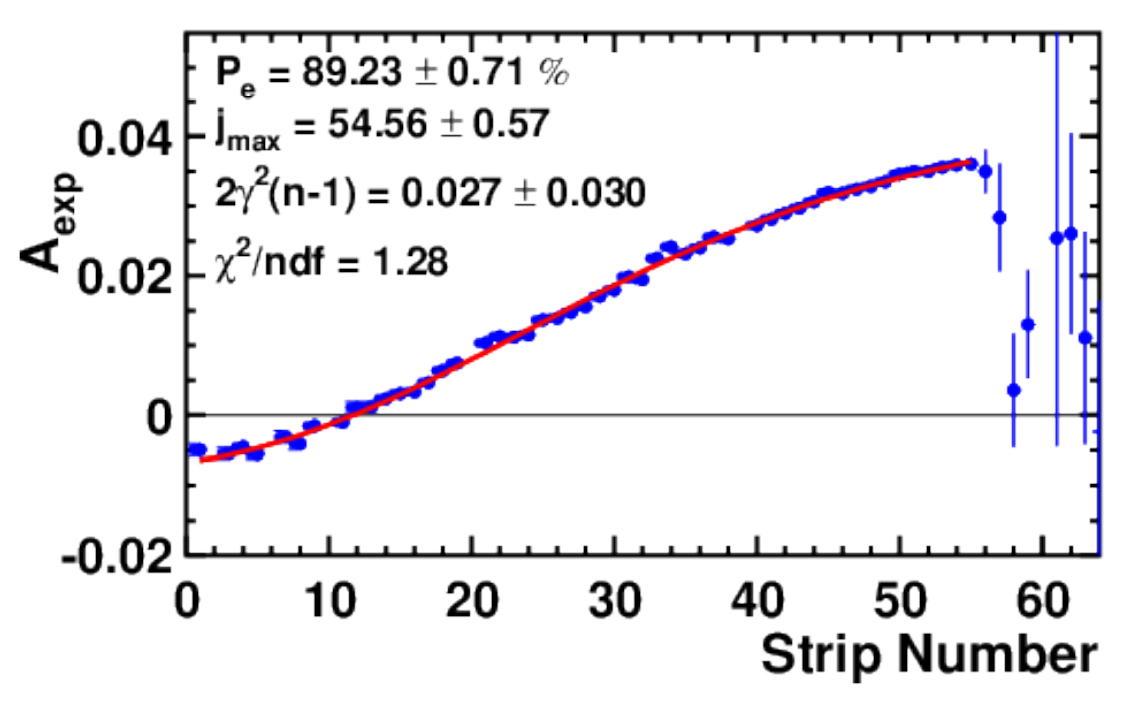}
   \includegraphics[scale=.4]{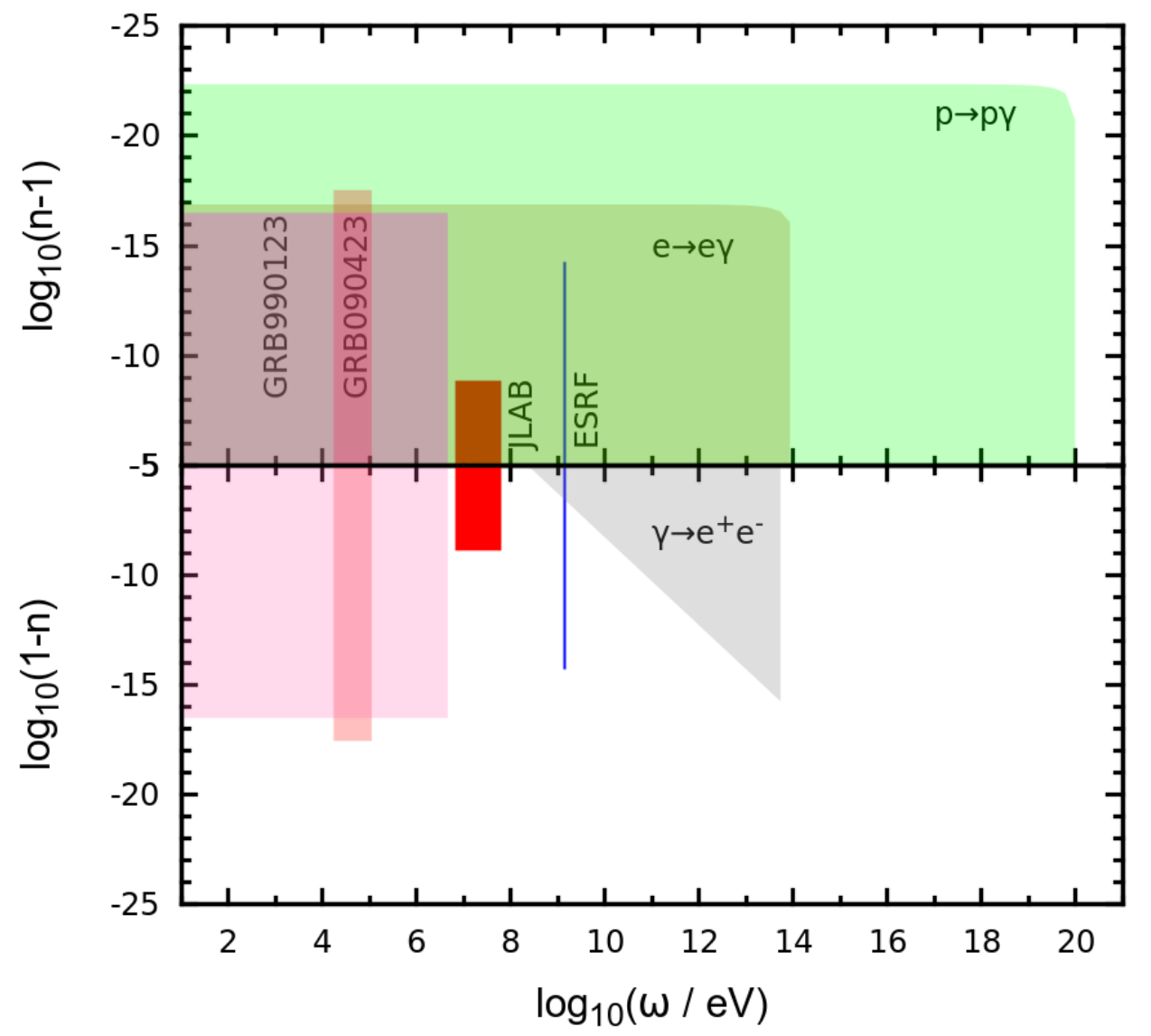}
\caption{\emph{Courtesy} Ref. \cite{[27]}.\\\textbf{(Left)} Measured Compton asymmetry (in blue) fit to the 3 free parameter-$\{P_e,  j_{max}, n\}$ function (in red) for a single run which lasts typically an hour long along with the corresponding fit values. \\\textbf{(Right)} Constraints on the speed of light \emph{w.r.t.} the energy of the photon obtained from Ch\'erenkov radiation, pair-production and GRBs along with the JLab measurement.}
\label{fig2}
\end{figure*}

The measured asymmetry is proportional to the product of polarization of photon and electrons involved in Compton scattering, $A^j_{exp} = P_eP_{\gamma}A^j_{th}$, where $j$ is the strip number. Strip number - $j$ gives us the energy of the scattered electron as the scattered electrons are analyzed by dipole - 3 and bend by an angle proportional to their energy. This allows us to write the asymmetry in terms of a dimensionless variable - $\rho = E_{\gamma}/E_{\gamma}^{max}$, where $E_{\gamma}$ is the energy of the scattered electron and $E^{max}_{\gamma}$ is the maximum allowed scattered electron energy dictated by the Compton edge.
\begin{eqnarray}
&A_{exp}(\rho)=\frac{2\pi r_e^2 a}{(d\sigma/d\rho)} \left( 1 - \rho(1+a)\right)\left[1-\frac{1}{(1-\rho(1-a))^2}\right]\label{eq3}\\
&\frac{d\sigma}{d\rho}=2\pi r_e^2 a\left[ \frac{\rho^2(1-a)^2}{1-\rho(1-a)}+1 + \left(\frac{1-\rho(1+a)}{1-\rho(1-a)}\right)^2 \right]\label{eq4}\\
&a = \frac{1}{1+4E_e^{beam}E_{\gamma}^0/m_e^2}\label{eq5}
\end{eqnarray}
where $r_e$ is the classical electron radius of $2.8$ fm and $E^0_{\gamma}$ is the incident photon energy. The asymmetry-$A_{exp}(\rho)$ can be written as a function of 3 free parameters \emph{viz.}, electron polarization-$P_e$, Compton edge-$j_{max}$, and refractive index of free space-$n$ \cite{[30]}.
\begin{eqnarray}
&\rho(n) \approx \rho\left[1+2\gamma^2f(x,\theta)\vec{\kappa}\cdot\hat{k}\right] \label{eq6}\\
&f(x,\theta) = \frac{(1+\gamma^2\theta^2)(1+x)^2 - (1+x^2+\gamma^2 \theta^2)^2}{(1+x+\gamma^2\theta^2)^2(1+x)^2}\label{eq7}
\end{eqnarray}
where, $x = 4\gamma \omega_0 \sin^{2}{(\frac{\theta_0^2}{2})}/m_e$, $\theta, \theta_0$ are the angle of scattered and incident electron respectively.
A global fit of the measured asymmetry to the 3 parameter function allows us to constrain $(n-1) < 6.8 \times 10^{-9}$ as shown in Figure \ref{fig2} (Left). Furthermore, this constraint on $(n-1)$ can be plotted \emph{w.r.t.} the energy of the photon together with constrains obtainable from:
\begin{itemize}
\item \underline{Ch\'erenkov radiation}: Using the highest energy particles (\emph{viz.} electrons, protons) observed.
\item \underline{Pair-production}: Using the highest energy photos observed.
\item and \underline{Gamma ray bursts (GRB)}: Using the time delay between the hard and soft part of the GRB spectra and the redshift associated to determine the distance (from the Earth) to the GRB.
\end{itemize}
as shown in Figure \ref{fig2} (Right). A sidereal modulation analysis of the refractive index extracted from each individual run was then performed to obtain the constraint $\sqrt{\kappa_X^2 + \kappa_Y^2} <  8.6 \times 10^{-10}$. The technique of measuring index of refraction from Compton scattering and the associated sidereal analysis is further detailed in Refs. \cite{[26],[27]}, along with the method of extracting constraints on $(n-1)$ \emph{w.r.t.} energy of the photon.

\section{CompEx-1b: Follow-up measurement in Hall-A of JLab}
\begin{figure*}[h]
\centering
   \includegraphics[scale=0.55]{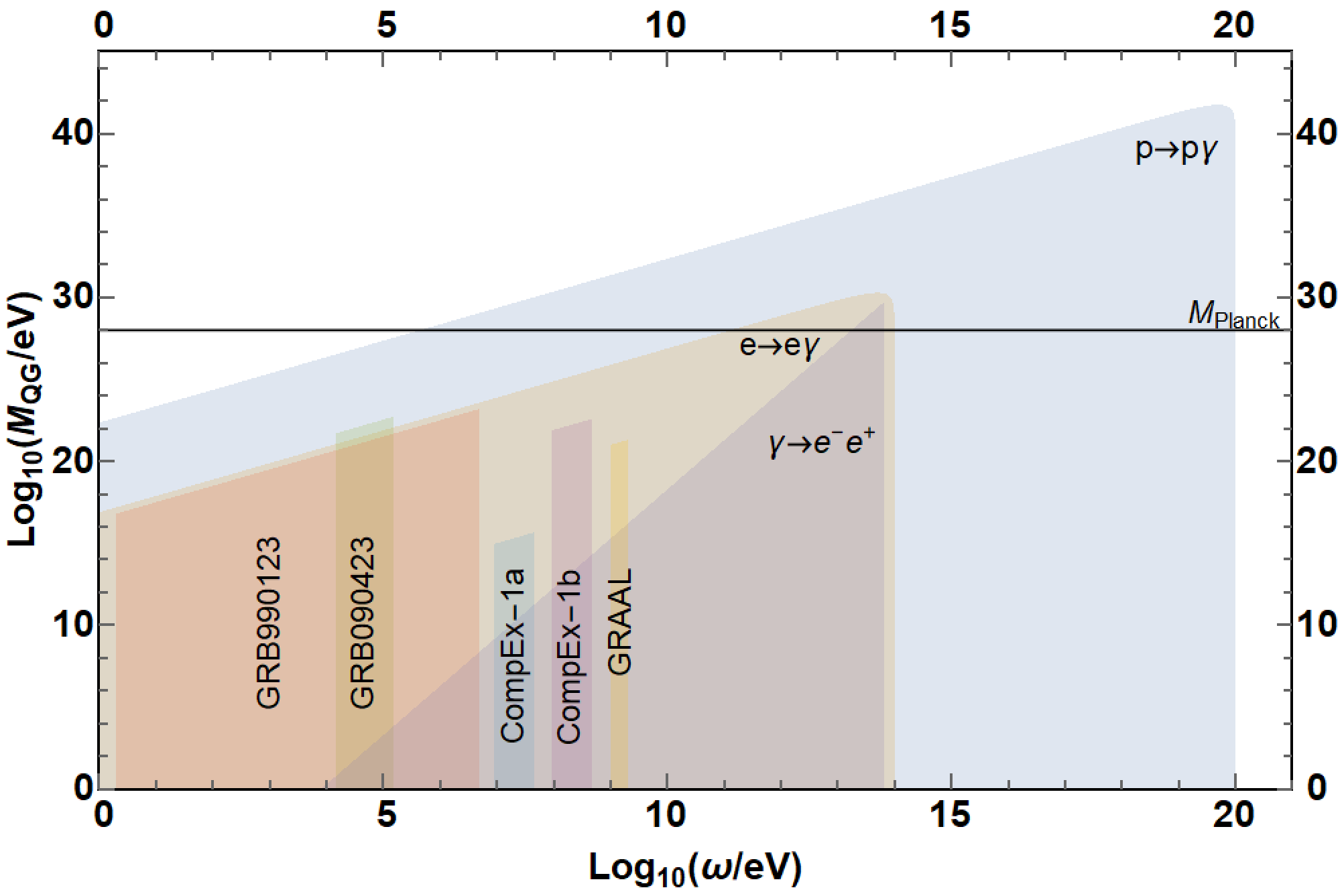}
   \caption[]{Constraints on the energy scale at which we can exclude LIV obtained from Ch\'erenkov radiation, pair-production and GRBs along with the JLab-CompEx-1a and GRAAL measurements. $M_{Planck}$ is the Planck mass of $\sim10^{28}$ eV and $\omega$ is the [scattered] photon energy.}
\label{fig3}
\end{figure*}

~~~The Hall-A Compton polarimeter was built and upgraded to provide a reliable, sub 1\%, measure of the electron polarization for the suite of parity violating (PV) experiments planned for 11 GeV program of JLab. Notably, for the MOLLER experiment which aims to measure the weak charge of the electron to sub 2.5\% precision, a measure of electron polarization precise to 0.4\% is required \cite{[28]}. The Hall-A Compton polarimeter looks similar to Hall-C Compton polarimeter, with the exception that the Fabry-P\'erot cavity has a quality factor as high as 3800 and is coupled to a frequency doubled 5 W, 1064 nm infrared laser \cite{[29]}. QWeak experiment used 1.16 GeV incident electrons whereas MOLLER shall use the full 11 GeV electron beam and a high beam current of up to 85$\mu$A. The $\ge10$-fold improvement in the power of the laser inside the Fabry-P\'erot cavity compensates for the linear reduction in the Compton scattering cross-section with increase in incident electron energy.

It is possible to convert the constraints in Fig \ref{fig2} (Right) to be expressed in terms of the mass scale at which we can exclude physics BSM. This can be done using first order LI violation (LIV) which modifies the speed of light linearly \emph{w.r.t.} the energy of the photon \cite{[32]}.
\begin{eqnarray}
n \sim 1 +\frac{E_{\gamma}}{M_{QG}} \label{eq8}
\end{eqnarray}
where $M_{QG}$ is thought be the scale of LIV. The constraints in Figure \ref{fig2} (Right) are reinterpreted in terms of Eq. \ref{eq8} and shown in Figure \ref{fig3}. When the measurement in Ref. \cite{[27]} using the Compton electron asymmetry is repeated at the JLab 12 GeV era, to calculate the increase in sensitivity, we can write the error in refractive index as a function of error in asymmetry with the help of Eqs. \ref{eq1}, \ref{eq3}-\ref{eq7}.
\begin{eqnarray}
\sigma_n = \frac{\sigma_{A_{exp}(\rho)}}{2\rho\gamma^2f(x,\theta)\sqrt{\frac{\partial^2A_{exp}(\rho)}{\partial\rho^2}}}
\label{eq9}
\end{eqnarray}
\begin{table}[h]
       \centering
        \begin{tabular}{c|l}
           $\chi^i$ & $\left( \chi^i_{1a}/\chi^i_{1b} \right)$ \\
           \hline
           $\sigma_{A_{exp}(\rho)}$ & 1/2\\
           $\rho$ & 9.48\\
           $\gamma^2$ & 89.92\\
           $f(x,\theta$) & 0.1\\
          $\partial^2 A_{exp}(\rho)/\partial \rho^2$ & 821\\
          \hline
          $\sigma_n$ & $1.4\times10^5$
        \end{tabular}
\caption{Amplification factor (per run $\sim 1h$) for the measurement of refractive index of free space, when comparing the measurement during 1.16 GeV QWeak (CompEx-1a) with measurement during 11 GeV MOLLER (CompEx-1b), arising from each parameter - $\chi^i$.}
\label{tab1}
\end{table}

The energy of scattered photons linearly scales with the incident electron energy. This results in an enhancement of around an order of magnitude when one goes from 1.16 GeV QWeak to 11 GeV MOLLER. The factor of 2 improvement in $\sigma_{A_{exp}(\rho)}$ mostly comes from the planned increase in the polarimetry precision from $\sim1\%$ during QWeak to $0.4\%$ for MOLLER. Given the high beam current and the length of the time MOLLER is planned to run, a follow-up measurement to Ref. \cite{[27]} is most ideally run as a parasitic experiment during the MOLLER beam time using Hall-A Compton polarimeter. Therefore, one could plug the appropriate operating conditions of MOLLER into Eq. \ref{eq9} and obtain an enhancement in sensitivity to refractive index, and this is reported in Table \ref{tab1}. Also, the factor of $1.4 \times 10^{5}$ reported for enhancement in $\sigma_n$ in Table \ref{tab1} does not take into account the fact that MOLLER plans to run for 344 days whereas QWeak only ran for 198 days. Taking the lengthier run time into consideration, we can expect an improvement by a factor of $1.9\times10^5$ over the constraints: $(n-1) < 1.4 \times 10^{-8}$ and $\sqrt{\kappa_X^2 + \kappa_Y^2} <  8.6 \times 10^{-10}$ (95$\%$ C.L.) resulting in projected sensitives of $(n-1) < 7.4 \times 10^{-14}$ and $\sqrt{\kappa_X^2 + \kappa_Y^2} <  4.5 \times 10^{-15}$ (95$\%$ C.L.) for CompEx-1b @ MOLLER in Hall-A. The region labeled as CompEx-1b in Figure \ref{fig3} shows this sensitivity in comparison to CompEx-1a @ QWeak in Hall-C and the current leading constraint provided by GRAAL: $\sqrt{\kappa_X^2 + \kappa_Y^2} <  1.6 \times 10^{-14}$ (95$\%$ C.L.) \cite{[34]}. Furthermore, a careful run-by-run analysis could lead to an additional improvement by a factor of $\sim2$ over and above the estimate shown in Figure \ref{fig3} for CompEx-1b @ MOLLER. Ergo an improvement over GRAAL constraint by a factor between 4 and 8 could be expected for the follow up measurement in CompEx-1b.

\section{CompEx-2: Future purview of search for Lorentz symmetry violation using Compton scattering}
\begin{table}[h!]
      \centering
        \begin{tabular}{l|c|c}
            Name & $E^{beam}_e$/GeV & $I^{beam}$/mA \\
            \hline
            Spring-8 & 8 & 100 \\
            ANL-APS & 7 & 100 \\
            PSI-SwissFEL & 5.8 & 20$\mu$ \\
            PETRA & 6 & 100 \\
	    SLAC-FACET & 20 & 100$\mu$ \\
            CERN-H8 & 100 & 1n \\
            \hline
            ILC & 250 & 8.7           
        \end{tabular}
\caption{Sources of high energy electrons along with electron beam current in mA (averaged over long periods \emph{i.e.} interpreted as a continuous current if the accelerator is operated in pulsed mode) and electron beam energy in GeV. \emph{Courtesy} Ref. \cite{[35]}}     
\label{tab2}
\end{table}
\begin{figure*}[h!]
\centering
   \includegraphics[scale=0.4]{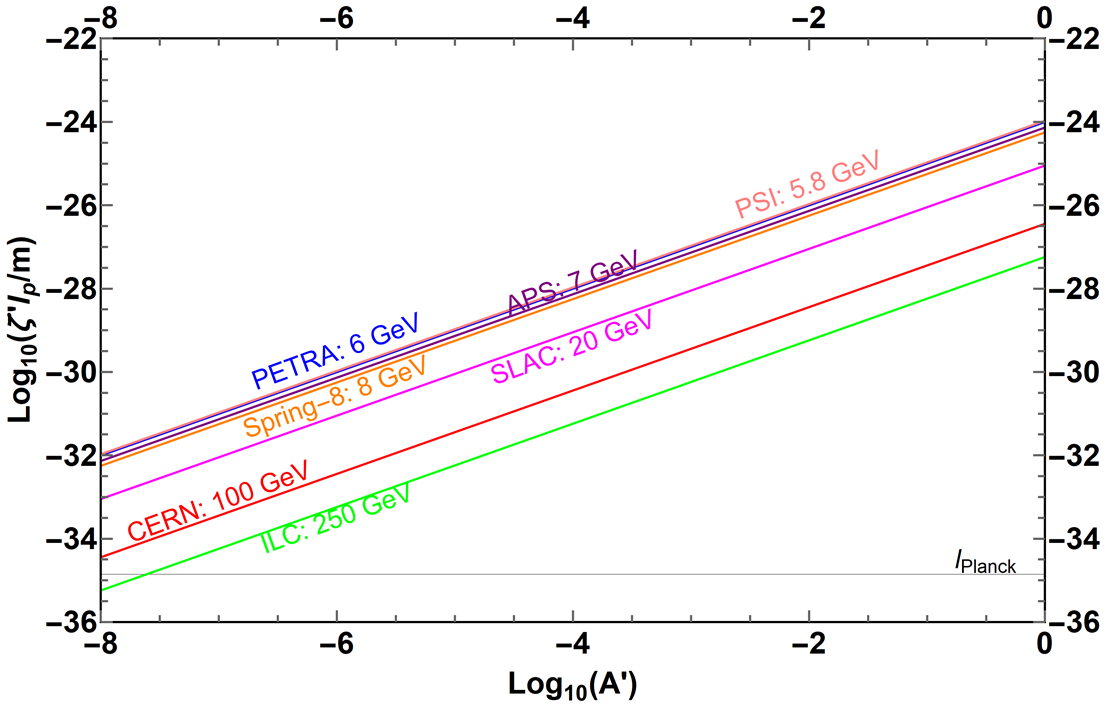}
   \includegraphics[scale=0.4]{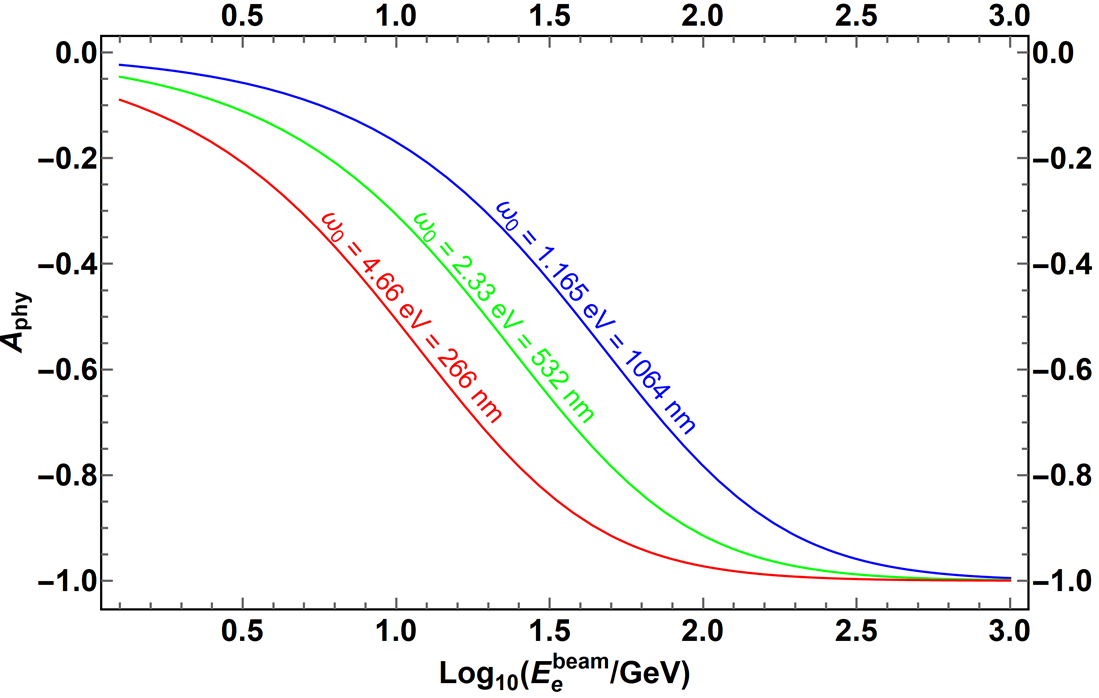}
   \caption[]{{\bf (Left)} Asymmetry induced due to vacuum birefringence expressed as a function of the length scale probed given by Eq. \ref{eq12}.\\ {\bf (Right)} Compton asymmetry at the Compton end point given by Eq. \ref{eq13} for varying incident photon energies.}
\label{fig4}
\end{figure*}

\begin{figure*}[h]
\centering
   \includegraphics[scale=0.85]{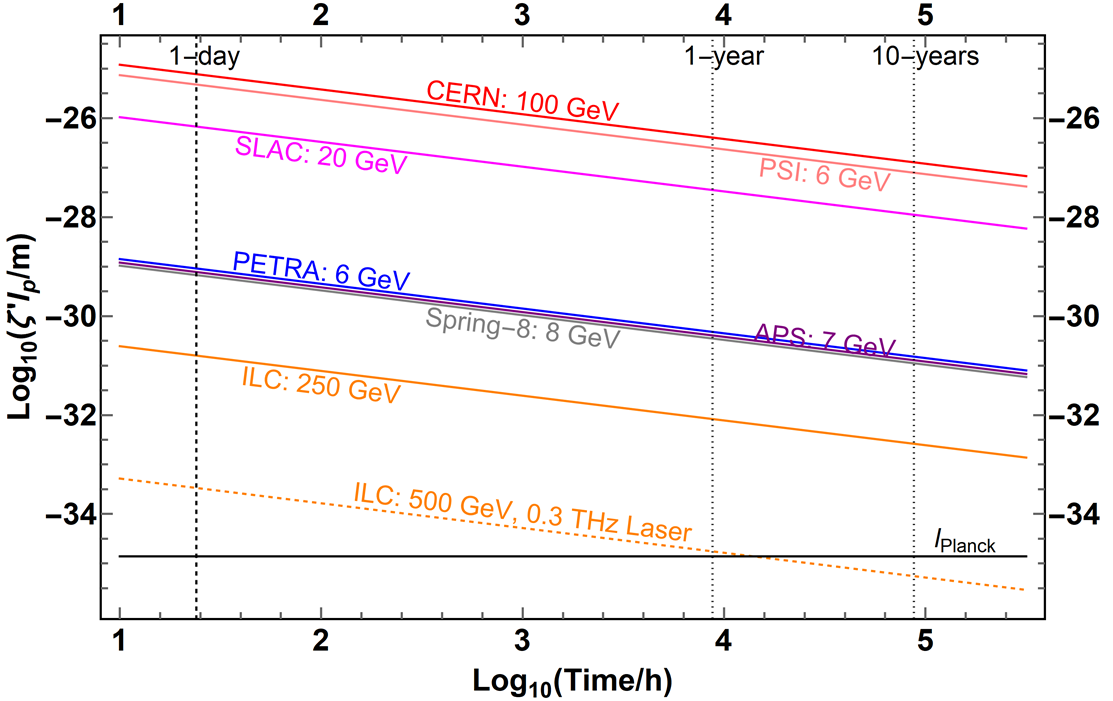}
   \caption[]{Reach of probing vacuum birefringence using Compton scattering at various electron accelerator facilities, where all the solid lines use $\omega_0 = $1.165 eV (1064 nm) @ 1.5 kW,  and dashed line uses $\omega_0 = $1 meV (0.3 THz) @ 1.5 kW.}
\label{fig5}
\end{figure*}
~~~Sidereal modulation of the value of refractive index exploits the anisotropy of speed of light as a signature for Lorentz symmetry violation. Figure \ref{fig3} shows us the extent of independence of speed of light \emph{w.r.t.} energy of the photon, both characteristic signals for breakage of LI. Birefringence is another means by which LI could be broken. Birefringence could occur due to the crystalline nature of space at Planck scales \cite{[extra]}. The location of Compton edge has a strong dependence on refractive index \cite{[31]};
\begin{equation}
j_{max}(n) - j_{max}(1) = \frac{32\gamma^6\omega_0Sin^4\left(\frac{\theta_0}{2}\right)}{(1+x)^4}\frac{\zeta'}{M_{QG}}
\label{eq10}
\end{equation}
where $j_{max}$ is location of the Compton edge in energy of the scattered photon (or Strip number), one of the fit parameters mentioned in Section 2, and $\zeta'$ is the dimensionless length parameter linking the photon energy to its momentum from Eq. \ref{eq8} as $(n-1) \sim (\zeta' k)/M_{QG}$. One could write $\zeta'$ in terms of Planck length, $l_p$ as $\zeta = \zeta'l_p$. 

In order to probe the vacuum birefringence, one could study the movement of Compton edge by changing the incident photon polarization state.
\begin{eqnarray}
A'&=&\frac{j_{max}^L - j_{max}^R}{j_{max}^L + j_{max}^R} \label{eq11}\\
A'&=&\frac{8\gamma^4\omega_0Sin^2\left(\frac{\theta_0}{2}\right)}{(1+x)^3}\frac{\zeta'}{M_{QG}} \label{eq12}
\end{eqnarray}
Eq. \ref{eq11} shows an asymmetry one could define using $j_{max}^{L(R)}$, the Compton edge while using left (right) circularly polarized incident light where the incident electron beam is unpolarized. Such asymmetries have obvious systematic advantages. Eq. \ref{eq10} shows us that small changes in refractive index are enhanced by a factor of up to $\gamma^6$ which allows us to compensate for the rather large value of energy scale ($\sim M_{QG}$) at which Lorentz symmetry could be violated. 

Figure \ref{fig4} (Left) plots Eq. \ref{eq11} which shows us the length scale at which we probe new physics as a function of asymmetry induced by vacuum birefringence. Compton scattering itself has an inherent asymmetry at the Compton edge given by \cite{[38]};
\begin{eqnarray}
A_{phys}&=&\frac{y^2-1}{y^2+1}\label{eq13}\\
y&=&\left( 1+\frac{4\omega_0 E^{beam}_e}{m_e^2}\right)^{-1}\label{eq14}
\end{eqnarray}
Figure \ref{fig4} (Right) plots Eq. \ref{eq13} for various incident photon energies. Compton asymmetry at $E_e^{beam}=1.1$ GeV, $\omega_0=2.33$ eV is $-0.04$ which was measured to the precision of $1\%$ in about an hour during Q-Weak \cite{[23]}. Therefore in an hour of data taking, we could have resolved an (with a C.L. of 1 $\sigma$) asymmetry as small as $4\times10^{-4}$. If no deviation is observed, then we could have excluded physics BSM that may have caused vacuum birefringence by looking at the $1$ GeV line on Figure \ref{fig4} (Left). This would then exclude such physics BSM at the level of $1.44 \times 10^{-26}$ m, some 9 orders of magnitude away from Planck scale. 

It is important to note that;
\begin{enumerate}
\item Compton asymmetry at the end point rises with increase in incident electron beam energy, according to Eq. \ref{eq13},
\item associated cross section reduces ($\sim$ linearly) with increase in incident electron beam energy, and
\item precision of measurement of the Compton asymmetry at the end point $\propto \sqrt{I^{beam}_e \cdot t}$ (where `$t$' is the time of data taking).
\end{enumerate}
We can normalize \#3 with Q-Weak beam time and beam current with associated precision of measurement in Ref. \cite{[23]} (also in the above para). By rolling in the above three effects, we can translate Figure \ref{fig4} (Left) to a length scale at which we can probe physics BSM caused by birefringent space as a function of length of time for which the experiment would need to be performed for. This is shown in Figure \ref{fig5}. Figure \ref{fig5} makes it evident that achieving Planck scale sensitivity using any of the currently available electron accelerators would take multiple decades of continuous data taking. Figure \ref{fig4} (Right) and Eq. \ref{eq13} shows that Compton asymmetry at the end point increases with incident electron beam energy and incident photon energy, therefore we could also calculate the sensitivity using incident photons of lower energy for which the asymmetry would be smaller. Figure \ref{fig5} includes a curve calculating the reach of this technique while using $500$ GeV electrons from International Linear Collider (ILC) Phase - V \cite{[35]} and Tera-Hertz laser in far IR regime. Tera-Hertz lasers with frequencies around 0.3 THz and Fabry-P\'erot cavities with high-Q factors for such high wavelengths have been demonstrated already \cite{[40]}.

\section{Conclusion}
~~~CompEx-1a analysis published in Refs. \cite{[26],[27]} serve as a proof of concept demonstrating the successful implementation of the analysis to measure the refractive index of free space using Compton asymmetry and the corresponding test of isotropy of speed of light as a signature of LIV. CompEx-1a measurement will be followed by the CompEx-1b program, using the cumulative data set of Hall-A Compton polarimeter collected during the series of PV experiments planned for 12 GeV era. CompEx-1b shall have a sensitivity better than the current leading constraints by a factor of around $\sim4 - 8$. CompEx-2 marks the next generation of Compton scattering based experiment in search of Lorentz symmetry violation and the future facility of 500 GeV ILC provides tantalizingly interesting possibilities with the ability to reach Planck scale sensitivities with about an year of data taking. 

\section{Acknowledgment}
~~~One of the authors (P. M.) would like to acknowledge the support from NSF-FCS grant \# 2015.0594 and the other (D. D.) would like to acknowledge support from DOE grant \# DE-FG02-07ER41528.

\end{document}